\documentclass[preprint,12pt]{elsarticle}

\usepackage{amssymb}
\usepackage{amsmath}
\usepackage{rotating}
\usepackage{graphicx}
\usepackage{color}

\usepackage[T1]{fontenc}
\usepackage{lineno,hyperref}

\modulolinenumbers[5]

\makeatletter
\def\ps@pprintTitle{%
 \let\@oddhead\@empty
 \let\@evenhead\@empty
 \def\@oddfoot{}%
 \let\@evenfoot\@oddfoot}
\makeatother

\begin{document}

\begin{frontmatter}

\title{Detecting correlations and triangular arbitrage opportunities in the Forex by means of multifractal detrended cross--correlations analysis}

\author[address3]{Robert G\k{e}barowski}

\author[address1]{Pawe{\l} O\'swi\k{e}cimka}
\author[address1,address2]{Marcin W\k{a}torek}

\author[address1,address2]{Stanis{\l}aw Dro\.zd\.z}

\address[address3]{Institute of Physics, Faculty of Materials Engineering and Physics, Cracow University of Technology, ul.~Podchor\k{a}\.zych~1, 30--084 Krak\'ow, Poland}
\address[address1]{Complex Systems Theory Department, Institute of Nuclear Physics, Polish Academy of Sciences,
ul.~Radzikowskiego 152, 31--342 Krak\'ow, Poland}
\address[address2]{Faculty of Computer Science and Telecommunications, Cracow University of Technology, ul.~Warszawska~24, 31--155 Krak\'ow, Poland}

\begin{abstract}
Multifractal detrended cross--correlation methodology is described and applied to Foreign exchange (Forex) market time series.
Fluctuations of high frequency exchange rates of eight major world currencies over 2010--2018 period are used 
to study cross--correlations. The study is motivated by fundamental questions in complex systems' response to significant
environmental changes and by potential applications in investment strategies, including detecting triangular arbitrage opportunities.
Dominant multiscale cross--correlations between the exchange rates are found to typically occur at smaller fluctuation levels.
However hierarchical organization of ties expressed in terms of dendrograms, with a novel application of the multiscale cross--correlation coefficient,
are more pronounced at large fluctuations. The cross--correlations are quantified to be stronger on average between those exchange rate pairs 
that are bound within triangular relations. Some pairs from outside triangular relations are however identified to be exceptionally strongly correlated
as compared to the average strength of triangular correlations.
This in particular applies to those exchange rates that involve Australian and New Zealand dollars and reflects their economic relations. 
Significant events with impact on the Forex are shown to induce triangular arbitrage opportunities which at the same time reduce cross--correlations 
on the smallest time scales and act destructively on the multiscale organization of correlations. In 2010--2018 such instances took place 
in connection with the Swiss National Bank intervention and the weakening of British pound sterling accompanying the initiation of Brexit procedure.
The methodology could be applicable to temporal and multiscale pattern detection in any time series.

\end{abstract}


\end{frontmatter}


\section{Introduction}
\label{intro}

Dynamics of complex systems, which are typically described by many degrees of freedom and a nonlinear internal structure, as well as their specific response to significant changes in the environment 
are within a research focus of many areas in fundamental and applied sciences, including mathematical, physical, biological and economic sciences \cite{Kwapien2012}. The present study of complexity in a system behavior 
is focused on the financial market with multicale signatures of its response to some major events influencing currency exchange rates.     

The foreign exchange financial market (known as the Forex or FX market) is a global market for currency trading in continuous operation 24 hours a day except for weekends (10pm UTC Sunday -- 10pm UTC Friday).
According to the recent foreign exchange statistics published by the Bank for International Settlements, the averaged daily trade volume exceeded 5 trillion US dollars in April 2016, 
with currencies mostly traded including US dollars, euros and Japanese yens (see --\cite{BIS2016}). This huge volume makes the Forex market the biggest in the world of finance and serves just as a single parameter
illustrating its enormous complexity \cite{Rickles2011}.

Seminal papers elucidating intricate complex dynamics of the foreign exchange market on short time scales include turbulent cascade approach \cite{Ghash1996},
motivated by hierarchical features in hydrodynamics of turbulence, as well as those referring to multi--affine analysis of typical currency exchange rates \cite{Vande1998a} and also their sparseness and roughness \cite{Vande1998b}.
The Forex market determines currency exchange rates between any traded pair of currencies through a complex and nowadays mostly machine driven automatic system of multiple type transactions among different kind of buyers and sellers round--the--clock.

Most recent works on uncovering patterns in foreign exchange markets include, but are not limited to, studies of lead--lag relationships \cite{Bosnarkov2020}, scaling relationships \cite{Boilard2018}, multifractality and efficiency issues \cite{Yang2019,Han2019}, partial correlations \cite{Bosnarkov2019} or quote spreads in high frequency--trading \cite{Cartea2019}.

The Forex market is highly liquid market due to its massive trading and asset volume. Typically it exhibits small daily fluctuations of currency exchange rates. 
Nevertheless, the value of one currency is determined relative to the value of the other currency through the exchange rate.
Any information or event having impact on one currency value will propagate through the Forex market by means of adjusting various other exchange rates, 
which could  be envisaged as a sort of a local (pairwise) currency interactions which eventually should influence globally all the currencies traded in the Forex market. Therefore we may expect that occasionally,
a significant fluctuation or even a temporally significant change in one exchange rate of any pair due to whatever reason will propagate to remaining pairs of currencies depending on the degree of cross--correlations between quotes 
for different currency pairs. Such influence should be particularly strong when comparing exchange rates of currencies with a common, the so--called base currency.
Any temporal discrepancy between exchange rates for such two pairs of currencies, with one currency in common, would offer immediately an opportunity to make a profit just by means of using this window of opportunity 
due to favorable exchange rates. It is the so--called triangular arbitrage opportunity \cite{Aiba2004,Fenn2009,Drozdz2010,Cui2019}. 

However in reality we deal with enormous amounts of data flowing at very high rates in the Forex market and elsewhere. 
Matching buy--sell transactions may typically occur during milliseconds via automated trading systems \cite{Buchanan2015}. Such enormous amounts of data have to be effectively analyzed for promising patterns down to a level of small fluctuations over small time scales. Big data, data mining, machine learning, artificial intelligence, an algorithmic high--frequency trading are in focus of quantitative finance \cite{Guida2019}.
Attempts to predict changes in financial time series from correlation--based deep learning has been quite promising \cite{Moews2019}. Recently, a novel predictive framework for forecasting future behavior of financial markets has been proposed by \cite{Ghosh2019}. Artificial intelligence may benefit a great deal from financial applications and social sciences alike \cite{Miller2019}. 
Hence financial market studies are of interest from fundamental and practical point of view for a wide community of scientists, engineers and professionals, bringing together many branches of mathematics, physics, economy
and computer sciences to address some challenging issues.

The main research problem addressed in this paper is the following: to what extent bivariate cross--correlations on the Forex market at various levels of fluctuations of exchange rates and time scales ranging from 
tens of seconds up to weeks may provide important information about a possibility to observe disparities in exchange rates, which may offer potential arbitrage opportunities. 

Our goal is to demonstrate prediction power of the so--called $q$--de\-trend\-ed cross--correlation coefficient stemming from the multifractal formalism when applied to historical time series of exchange rates for a set of currencies.
We will investigate in details sensitivity of various statistical properties of small and large fluctuations using natural scalability of our cross--correlation measure and follow closely the way how information and events related to financial markets may build arbitrage opportunities among a set of currency exchange rates.

We would like to emphasize that our method based on detrended cross--correlation analysis is quite novel and only recently a plethora of applications started to emerge across many fields of nonlinear correlations studies, including meteorological data \cite{Chenhua2018}, electricity spot market \cite{Fan2015}, effects of weather on agricultural market \cite{Cao2016}, stock markets \cite{Zhao2018}, cryptocurrency markets \cite{Drozdz2019}, electroencephalography (EEG) signals \cite{Chen2018}, electrocardiography (ECG) and arterial blood pressure \cite{Ghosh2019b} as well as air pollution \cite{Shen2017,Wang2017}. Such wide interest across different fields of research in application of detrended cross--correlation analysis to nonlinear time series studies serves as an additional strong motivation for elucidating such analysis in terms of its potential and limitations.

The paper is organized as follows. First, we discuss the Forex data used in the present study and our methodology for the financial times series of logarithmic returns for exchange rates. Next, we discuss a degree of triangular
arbitrage opportunity in a form of a convenient coefficient derived from suitable exchange rates. Then, we describe fundamental concepts of our statistical methods, which stem from multifractal formalism. 
We define $q$--dependent de\-trend\-ed coefficient capturing cross--correlations of two detrended time series. The results section comprises global behavior of currency rates, logarithmic return rates statistics with the focus 
on large fluctuations, discussion of hierarchy among currency exchange rates and the role of abrupt cross--correlation changes and large fluctuations in detecting arbitrage opportunities.
Finally, we summarize and draw some general conclusions.

In view of above mentioned interdisciplinary research by other authors \cite{Chenhua2018,Cao2016,Zhao2018,Chen2018,Ghosh2019b,Shen2017,Wang2017}, through the conclusions from our present work, we would like also to support advantages of multifractal detrended cross--correlation method and its wide applications to study any time series with nonlinear correlations, not only in the foreign exchange market but also across other fields of pure and applied sciences.

\section{Data and financial time series methodology}

The data used in the present study has been obtained from the Dukascopy Swiss Banking Group \cite{data}.

The data set (see the Appendix) comprises foreign exchange rates for the period 2010 -- 2018  between pairs of 8 major currencies.

For the purpose of this study we consider the arithmetic average out of the bid and ask price for each exchange rate:

\begin{equation}
R(t) = \frac{ R_{\textrm{ask}}(t) + R_{\textrm{bid}}(t) }{2}.
\end{equation}

Then we consider the following time series of logarithmic returns of such exchange rates for each pair of currencies:

\begin{equation}
r(t) = \log (R(t + \Delta t)) - \log (R(t)),
\end{equation}

which are then processed as described in the Appendix.

\section{Triangular arbitrage in the Forex market}

Let us first consider a model situation where we can instantly carry out a sequence of transactions
with exchange rates, which all of them are known for a given time instance $t$. In our example a subset of all exchange rates for 3 currencies (EUR, USD, CHF), namely 3 quotes { EUR/CHF, EUR/USD, USD/CHF } is considered. 
Let us assume that a trader holds initially euros (EUR). One possibility to use arbitrage opportunity, would be to do a sequence of transformations including those 3 currencies \cite{Aiba2004,Fenn2009}:

\begin{equation}
\textrm{EUR}  \rightarrow \textrm{USD}  \rightarrow \textrm{CHF} \rightarrow \textrm{EUR}.
\end{equation}

In this case the rate product (we buy USD with EUR, then buy CHF with USD and finally sell CHF in return for EUR):

\begin{align}
\alpha_1 (t)  & = R_{\textrm{bid}}(\textrm{\scriptsize EUR/USD}, t) \cdot  R_{\textrm{bid}}(\textrm{\scriptsize USD/CHF}, t) \nonumber \\
              & \qquad \times \frac{1}{R_{\textrm{ask}}(\textrm{\scriptsize EUR/CHF}, t)} - 1
\label{a1}
\end{align}

If such a rate product $\alpha_1 > 0$ then the arbitrage is possible, provided all transactions can be completed with such instant rates. That is we could end up with more currency EUR than we had initially. 
In practice however this is difficult on real markets and in fact after the first leg of such multiple transactions, remaining trades would not be possible to complete or the price will be changed by the time they 
will be completed. 

Another alternative possible minimal exchange currency path in our case  (again assuming all the transactions are done in the same time instance or with the frozen exchange rates) would be the following

\begin{equation}
\textrm{EUR}  \rightarrow \textrm{CHF}  \rightarrow \textrm{USD} \rightarrow \textrm{EUR}.
\end{equation}

Its value in terms of the product of exchange rates will be the following (replicated by means of selling EUR for CHF, then selling CHF for USD and finally buying back EUR for USD):

\begin{align}
\alpha_2 (t)  & = \frac{1}{R_{\textrm{ask}}(\textrm{\scriptsize EUR/CHF}, t)} \cdot \frac{1}{R_{\textrm{ask}}(\textrm{\scriptsize USD/CHF}, t)} \nonumber \\
              & \qquad \times R_{\textrm{bid}}(\textrm{\scriptsize EUR/USD}, t) - 1
\label{a2}
\end{align}

Including more currencies in our basket we could consider longer exchange paths allowing to use more factors (appropriate exchange rates) in our $\alpha$ rate product coefficient. Again its value compared to one would 
indicate a theoretical possibility of executing arbitrage opportunity.

\section{Multifractal statistical methodology}

Let us consider multiple time series of exchange rates re\-cord\-ed simultaneously. We are interested in a level of cross--correlation between a pair of non-stationary time series, ${ x(i) }$ and ${ y (i) }$, 
where $i=1,2, \ldots, N$ corresponds to subsequent instances of time when the signals were recorded ($t_1 < t_2 < \ldots < t_N$). Both time series are synchronized in time and have the same number $N$ of data points.

Following the idea of a new cross--correlation coefficient defined in terms of detrended fluctuation analysis (DFA) and detrended cross--correlation analysis (DCCA)--\cite{Podobnik2008},
which has been put forward in \cite{Zebende2011}, we use in the present study a multifractal detrended cross--correlation analysis (MFCCA) \cite{Oswiecimka2014} 
with $q$--depended cross--correlation coefficient. The cross--correlations of stock markets have been also investigated with a time--delay variant of DCCA
method \cite{Lin2012}. Multiscale multifractal detrended cross--correlation analysis (MSMF--DXA) has been proposed and subsequently 
employed to study dynamics of interactions in the stock market \cite{Lin2014}. Other methods, including weighted multifractal analysis of financial time series 
\cite{Xiong2017} and multiscale properties of time series based on the segmentation \cite{Xu2018} allow for multifractal and multiscale nonlinear effects investigations.

Multifractal cross--correlations can also be studied using Multifractal Detrending Moving--Average Cross--Correlation Analysis (MFXDMA) as it has been done by \cite{Jiang2011} 
and within the wavelet formalism by making use of the Multifractal Cross Wavelet Transform (MF-X-WT) analysis performed by \cite{Jiang2017}. 
The approach adopted in present study has been introduced by \cite{Kwapien2015}. In what follows, we briefly state main points of this approach.

\subsection{Detrended cross--correlation methodology}

We define a new time series $X(k)$ of partial sums of the original time series elements $x(i)$. In an analogous way, the second time series of interest, $Y(k)$, is obtained out of the original time series $y(i)$.
Depending on the nature of the signal we may expect in time series trends and seasonal periodicities. In order to remove such trends at various time scales $s$ (if data points are collected at the same
time intervals $\Delta t$, the scale number $s$ corresponds to the time) we subdivide both series of length $N$ into a number of $M(s) = \textrm{int}(N/s)$ non--overlapping intervals, each of length $s$ ($s < N/5$),
starting from one end of a time series. For a given time scale $s$ we may repeat the partitioning procedure from the other end of the time series, thus obtaing in total $2 M(s)$ time intervals, each of which will 
contain $s$ data points. For each of the interval $k$ ($k = 0, \ldots, 2M(s)-1$), we find a polynomial of degree $m$, $P^{(m)}_{(s,k)}$, for a given time scale $s$ 
(eqivalent to the number $s$ of used data point to perform the fit). In general it could be a polynomial of any finite degree, depending on the nature of the signal. 
For the purpose of data analysis in this paper we use polynomials of degree $m=2$ unless otherwise stated \cite{Oswiecimka2014}.

Now we are ready to construct an estimate of the covariance of both newly derived time series, each of which have those polynomial trends removed interval by interval:

\begin{align}
C_{XY}^{(s,k)} &  =  \frac{1}{s} \sum_{j=1}^{s} \ \left [ X \bigl ( s k + j \bigr )  -  P^{(m)}_{(s,k)} \left [ X \bigl ( s k + j \bigr ) \right ] \right ]  \nonumber \\
               &  \qquad \quad \times \left [ Y \bigl ( s k + j \bigr )  -  Q^{(m)}_{(s,k)} \left [ Y \bigl ( s k + j \bigr ) \right ] \right ]
\label{cxy}
\end{align}

If both time series are the same, we obtain an estimate of the detrended variance in the $k$--th interval from partitioning with the scale $s$. Constructed estimates of the covariance allow us to define
a family of fluctuation functions $F_{XY}^{(q)}(s)$ of the order $q \ne 0$ for a given choice of the time scale (number of data points in the interval) $s$
\begin{equation}
F_{XY}^{(q)} (s) = \frac{1}{2 M(s)} \sum_{k=0}^{2M(s)-1} \ \textrm{sign} \bigl [ C_{XY}(s,k) \bigr ] \ \bigl | C_{XY}(s,k) \bigr |^{q/2}
\label{fqs} 
\end{equation}
For $q=0$ we use a formula for the computation of the fluctuation functions, which essentially follows from de l'Hospital rule \cite{Kantelhardt2002,Oswiecimka2014}.

The definition of the family of fluctuation functions given by Eq.~(\ref{fqs}) could be applied in principle also for the case when both time series are the same. 
In such a case we obtain, as it should be, the $q$--th order fluctuation function, which follows from the multifractal detrended fluctuation analysis (MFDFA) for a single time series \cite{Kantelhardt2002}. 
Multifractal cross--correlations for time series  should manifest themselves as a power--law behavior of the fluctuation functions  $F_{XY}^{(q)}$ over a range of time scales $s$:

\begin{equation}
\bigl [F_{XY}^{(q)} \bigr ]^{1/q}=F_{XY}(q,s) \sim s ^{\lambda(q)} 
\label{lambda}
\end{equation}

The generalized $q$--dependent exponent, $\lambda(q)$, provides the insight into multifractal properties of the cross--correlations. In the limiting case of $\lambda(q)=const$, 
that is in the case of the exponent being independent on $q$, we may infer the monofractal type of cross--correlations between the two time series $x(i)$ and $y(i)$. Again, in the case of $x(i)\equiv y(i)$,
the asymptotic behavior of the fluctuation function, as defined by Eq.~(\ref{fqs}), takes a form which is familiar from the MFDFA approach:
\begin{equation}
\bigl [F_{XX}^{(q)} \bigr ]^{1/q}=F_{XX}(q,s) \sim s^{h_{X}(q)},
\label{lamfdfa}
\end{equation}
where $h(q)$ is a generalized Hurst exponent  \cite{Kantelhardt2002,Drozdz2009,Drozdz2015,Grech2016,Klamut2018}.

\subsection{Detrended cross--correlation $q$--coefficient}

The family of fluctuation functions of order $q$, defined by Eq.~(\ref{fqs}), convey the information about fluctuations of different order in magnitude over a range of time scales.
Therefore one can define a $q$--dependent detrended cross--correlation ($q$DCCA) coefficient using a family of such fluctuation functions \cite{Kwapien2015}:
\begin{equation}
\rho_q(s) = \frac{ F_{XY}^{(q)}(s) }{ \sqrt{ F_{XX}^{(q)}(s) \ F_{YY}^{(q)}(s) } }.
\label{rhoq}
\end{equation}
The coefficient $\rho_q(s)$ allows to measure a degree of the cross--correlations between two time series $x_i$, $y_i$. It has been proven that the $q$--dependent cross--correlation coefficient is bounded:
$-1 \le \rho_q(s) \le 1$, for $q > 0$ \cite{Kwapien2015}. 

The definition of the cross--correlation given by Eq.~(\ref{rhoq}) can be interpreted as a generalization of the Pearson coefficient \cite{Pearson1895,Rodgers1988}. 
Advantage of the coefficient  $\rho_q(s)$ over the conventional Pearson index has already been studied through numerical methods in different contexts by several authors (e.g. see \cite{Jiang2018} and references therein).
The parameter $q$ helps to identify the range of detrended fluctuation amplitudes corresponding to the most significant correlations for these two time series~\cite{Kwapien2015}. 

It has been established that for $q >2$ large fluctuations mainly contribute to $\rho_q(s)$, whereas for $q < 2$ the dominant contribution is due to small and medium size fluctuations \cite{Kwapien2015,Zhao2018}.
Hence by choosing a range of values in $q$ we may filter out correlation coefficient for either small or large fluctuations.

\section{Analysis and results}

Based on the methodology described, we will investigate multiscale properties for cross--correlations among time series corresponding to currency exchange rates for the whole period 2010--2018 as well
as for some sub--periods.

We use MATLAB desktop environment for the numerical implementation of our methodology. Although we have quite large data sets (see the Appendix), the implementation itself is quite straigtforward and all the computations can be accomplished on a modern PC class computer. We also apply standard methods of MATLAB source codes validation and surrogate data checking against artefacts or robustness of nonlinear correlations within our data sets \cite{Kwapien2015}.

Although in our study we focus on different signatures and statistical properties of multivariate time series with respect to triangular arbitrage, one may envisage a broader picture of such analysis, whereby one would like to uncover a specific kind of cross--correlations in these time series which would help us to detect underlying interconnections useful for the system behavior prediction in future.

\subsection{Logarithmic returns and the inverse cubic law}

In this subsection we will present a general picture for the financial time series dynamics with an emphasis on events which have an impact on the Forex market.  

\subsubsection{A currency index}

We define a currency index $CI_a$ for the currency $a$ as an average out of cumulative log--returns for all exchange rates that have $a$ as a base currency.

The $CI_a$ could be cast in the following form:
\begin{equation}
CI_a (t) = 1 +  \frac{1}{N-1} \sum_{b \ne a} r(b/a, t),
\label{ci}
\end{equation}
where $N=8$ in the present case and $r(b/a, t)$ denotes time--dependent log--return for exchange rate $b/a$ with 
label \\
$b \in \{ AUD, CAD, CHF, EUR, GBP, JPY, NZD, USD \}$ \\
such that it is different from the label $a$.

\begin{figure}[h!]
\centering
\includegraphics[scale=0.5,keepaspectratio=true]{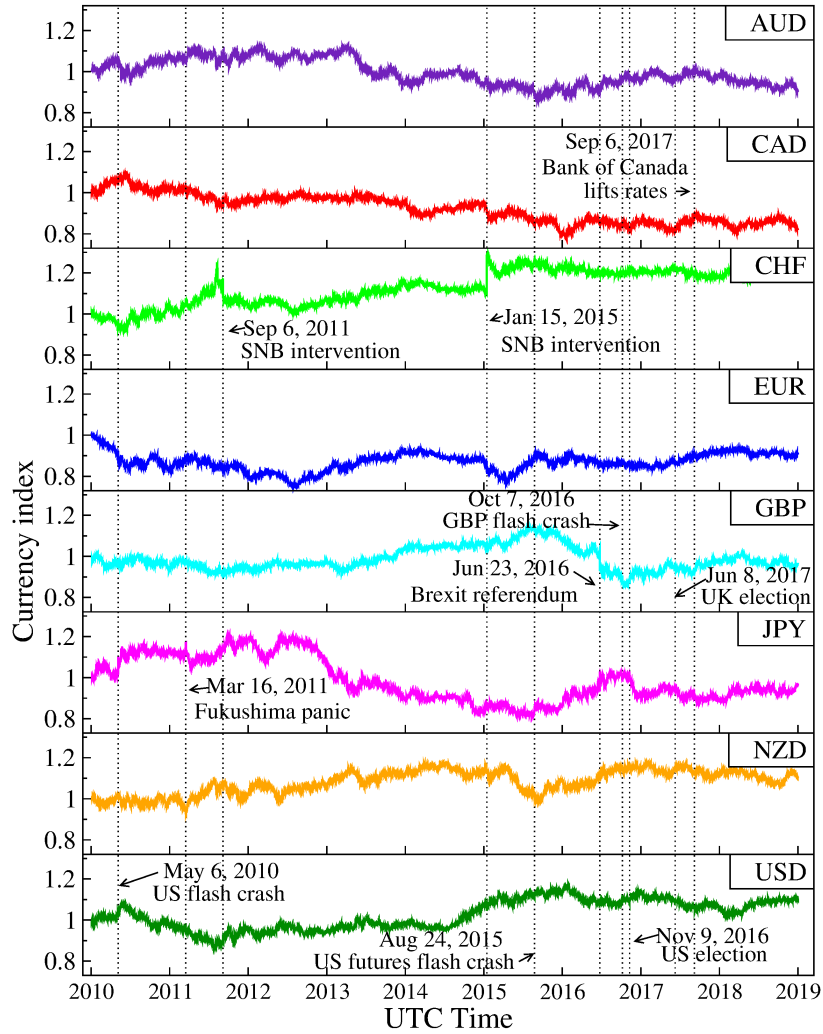}
\caption{(Color online) Currency indexes as defined by Eq.~(\ref{ci}) between years 2010 and 2018. For a reference, some important global political and economic events have been indicated over the time scale.
        }
\label{fx}
\end{figure}

The currency index given by Eq.~(\ref{ci}), indicates performance of any given currency in terms of exchange rates to other currencies in our data sets. With such a single averaged characteristics, 
one may have a general overview of a global temporal behavior and performance of any currency in the Forex market.

In Fig.~\ref{fx} currency indexes time variation are shown for all the currencies in our data set for
the considered period 2010 -- 2018.  For this plot we take logarithmic returns arising from average bid and ask exchange rates. In the figure we have indicated some political or economic events on the time scale
(with dotted vertical line) which in principle could have impact on the Forex market performance during that period of time. These labels may serve as an intuitive explanation of features observed on the curves
related to the currency indexes. We have included some sudden events: the US stock market flash crash on the 6th of May 2010, Fukushima Daiichi (Japan) nuclear disaster unfolding during several days in the aftermath
of the earthquake which took place on the 11th of March 2011, the Swiss National Bank (SNB) interventions
(on the 6th September 2011 and 15th January 2015), the US futures flash crash on 24th August 2015, the Brexit referendum on the 23rd of June 2016, GBP pound sterling's flash crash on the 7th of October 2016 
(GBP/USD exchange rate briefly dropped overnight by 6 per cent), the US Presidential elections (on the 9th November 2016) and the general elections in the UK (on the 8th of January 2017), 
and increase of overnight rate target to 1 per cent by the Bank of Canada (on the 6th September 2017).

In general, one observes significant variations of considered currency indexes over the period of 8 years. 
Worth noting in this timeline are Swiss National Bank (SNB) interventions in 2011 and in 2015 as well as the Brexit referendum in 2016. In the following, we will explore in some more details statistical properties of 
the Forex market data in the vicinity of these events. We will discuss to what extent our proposed statistical analysis corroborates these features, when looking from the hindsight with the help of 
historical data from the Forex market.

\subsubsection{Inverse cubic tails of absolute return distributions} 

Another interesting feature of the Forex market is related to the cumulative distribution of absolute logarithmic returns:
\begin{equation}
P(X > |r|) \sim |r_{\Delta t}|^{\gamma}
\label{P}
\end{equation}

\begin{figure}[h!]
\centering
\includegraphics[scale=0.45,keepaspectratio=true]{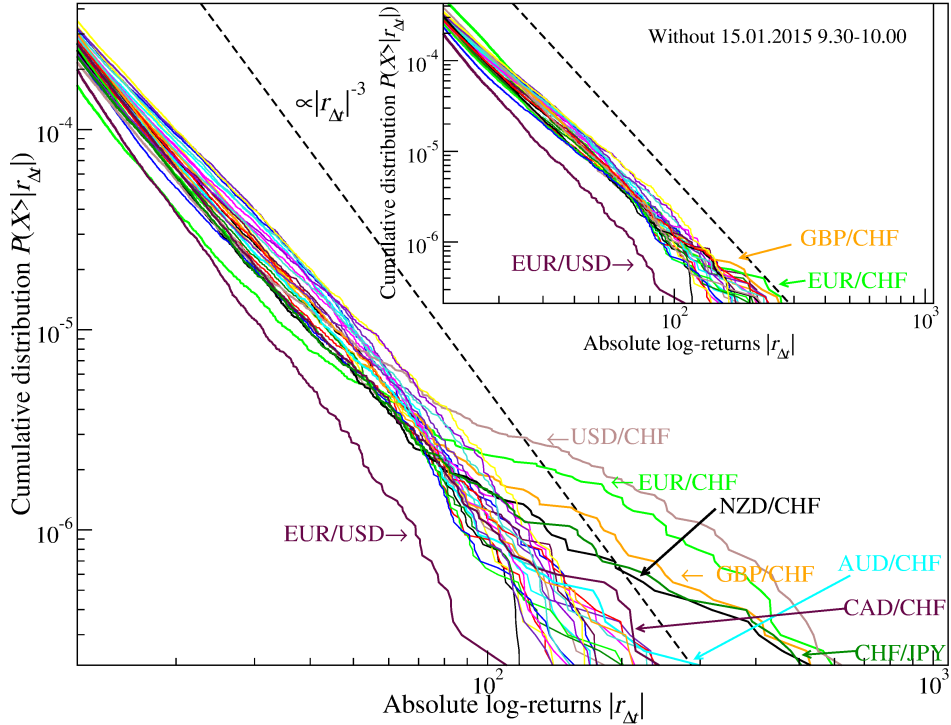} 
\caption{(Color online) The double log plot showing tails of the cumulative distribution of normalized absolute log--returns $|r_{\Delta t}|$ for currency exchange rates in the whole period 2010--2018. Black dashed line shows the slope $\alpha = -3$ which corresponds to the approximate inverse cubic power-law. The inset shows these distributions when a short period of an extreme volatility in currency exchange rates has been removed. This removed period (a half an hour) corresponds to the wake of the SNB intervention on the 15th January 2015.}
\label{inv3}
\end{figure}

In  Fig.~\ref{inv3} we show such cumulative distributions for exchange rate absolute normalized returns among all major currencies considered in the present study. Each solid line of different color demonstrates 
the tail behavior for the corresponding currency. In this double--log plot the black dashed line corresponds to the slope of $\gamma_0 = -3$, which is the so--called inverse cubic 
power-law~\cite{Gopikrishnan1998,Drozdz2003,Drozdz2007}. The least ''fat tail'' corresponds to returns of the exchange rates EUR/USD, which typically form the most liquid pair of currencies in the Forex market.
Significant deviations from the inverse--cubic law are clearly  present for large absolute log--returns ($|r_{\Delta t}| > 10^2$ for exchange rates involving CHF (Swiss franc)).
The tails for the following exchange rates USD/CHF, EUR/CHF, GBP/CHF, NZD/CHF are significantly ``fatter'' than it would follow from the inverse cubic behavior 

We note that all currency exchange rates with the Swiss franc (CHF) as base currency yield higher probability of larger absolute logarithmic returns than other exchange rates. 
These outliers could be attributed to two instances of the SNB interventions (in 2011 and 2015). In order to demonstrate the origin of these deviations, we remove from our data sets a period of a half an hour
in the morning on the 15th of January, 2015, when a significant volatility of currencies exchange rates has been observed in the wake of the SNB intervention \cite{Yang2019}. 
The resultant tails of the probability distributions are shown in the inset of Fig.~\ref{inv3}. Hence without this single, short--termed event on the market, the tails of the distributions approximately follow the inverse--cubic behavior.

\subsection{Multifractality and scaling behavior of cross--correlation fluctuation functions}

We already know, that the outliers of the cumulative distributions document an increased level of larger fluctuations in absolute log--returns. This means also a higher chance to encounter fluctuations yielding larger returns.
The question arises to what extent these fluctuations are cross--correlated among exchange rates. Such cross--correlations at least between two exchange rate time series would offer a potential opportunity of triangular arbitrage.

\begin{figure}[h!]
\centering
\includegraphics[scale=0.45,keepaspectratio=true]{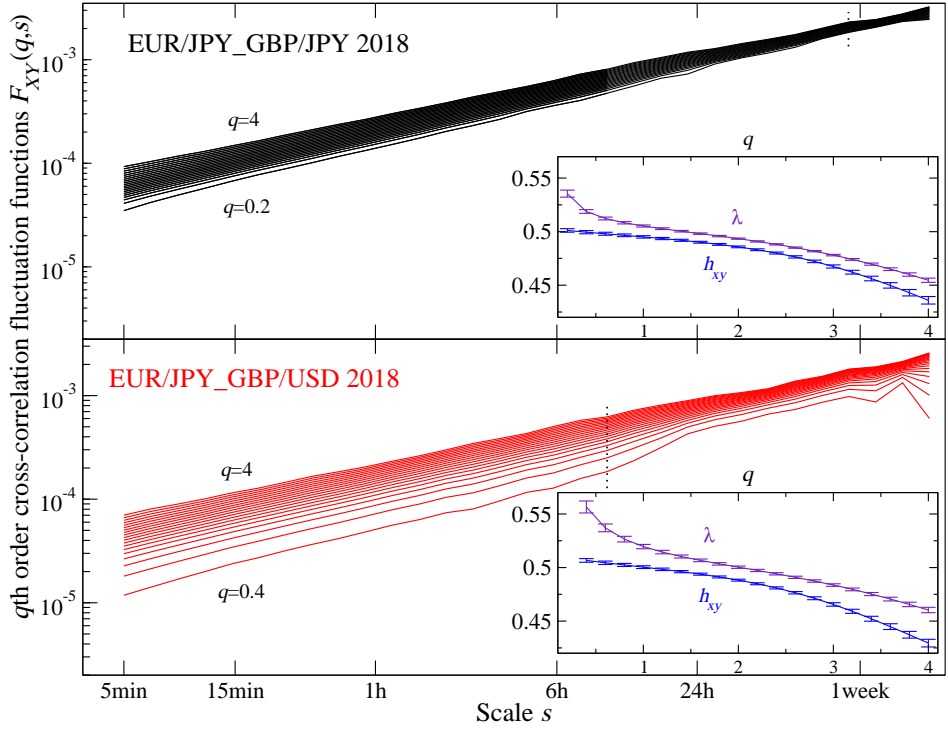} 
\caption{(Color online) An example of the results for $q$-th order cross--correlation fluctuation functions
$F_{XY}(q,s)$ for EUR/JPY with GBP/JPY (the top panel) and for EUR/JPY with GBP/USD logarithmic return exchange rates (the bottom panel). The graph illustrates the cross-correlation fluctuation functions scaling over the range of time scales $s$  from 5 minutes up to 2 weeks for different $q$--coefficients. The insets show scaling exponent $\lambda(q)$ dependence on the coefficient $q$ and the average $h_{XY}(q)$ of the generalized Hurst exponents obtained for each of the time series individually.}
\label{Fqs}
\end{figure}

Let us begin by investigating an example of multifractal cross--correlations for time series in terms of cross--correlation fluctuation functions $F_{XY}(s,q)$. Fig.~\ref{Fqs} shows the results 
for a cross--correlation fluctuation functions between EUR/JPY with GBP/JPY and EUR/JPY with GBP/USD for exchange rates in a single, 2018 year.

In Fig.~\ref{Fqs} (the main panel), each line in the bunch of lines corresponds
to the fluctuation function $F_{XY}(q_i,s)$ with $0.2 (0.4) \le q_i \le 4$. We verify indeed that the scaling according to Eq.~(\ref{lambda}) is valid in the range $0.8 \le q \le 4$ that is the slope of 
the line in the double--log plot is approximately independent on the scale $s$ in the range shown. We may also observe that the fluctuation functions depend on the coefficient $q$ for $1 \lessapprox q  \lessapprox 4$
though in this case that dependence is rather weak. Additionally, for the reference both insets in Fig.~\ref{Fqs} show the average $h_{XY}(q)$ of generalized Hurst exponents for each individual time series. 
This average is defined as follows \cite{Zhou2008}:

\begin{equation}
 h_{XY}(q) = \frac{h_X(q) + h_Y(q)}{2}
\label{hxy} 
\end{equation}

It is worth noting that in this shown example, the scaling of the fluctuation functions for the case of the triangular relation among two exchange rates (the top panel in Fig.~\ref{Fqs}) is much better preserved
for small $q$ parameters than this is the case for the exchange rates which are outside the triangular relation (the bottom panel).

\subsection{Detrended cross--correlation q--coefficient in the Forex market}

\begin{figure}[h!]
\centering
\includegraphics[scale=0.45,keepaspectratio=true]{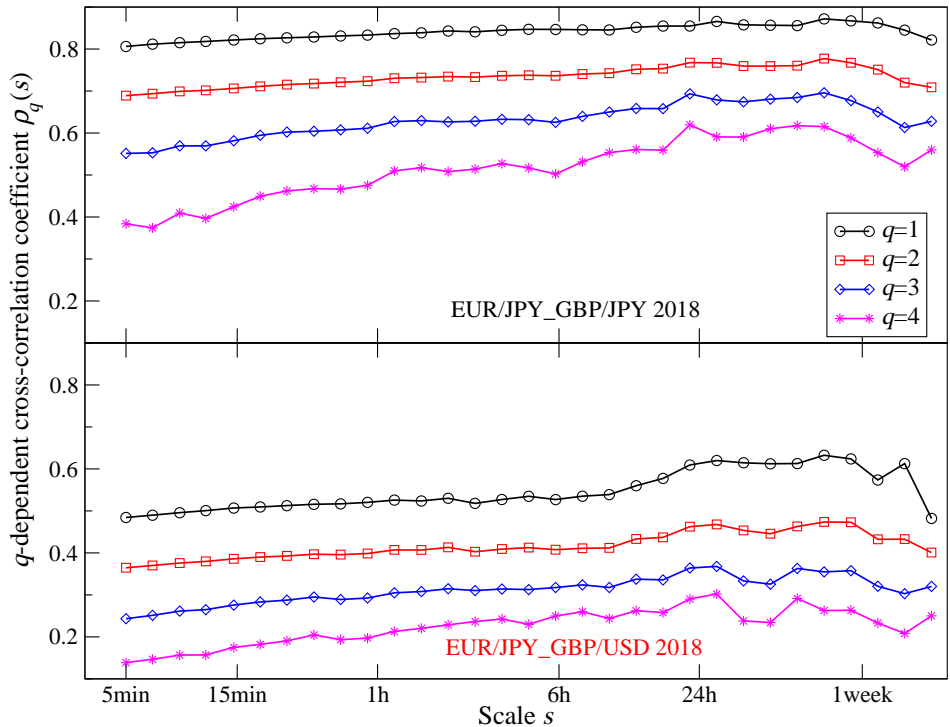} 
\caption{(Color online) An example of the results for $\rho_q(s)$ cross--correlations for EUR/JPY with GBP/JPY (the top panel) and for EUR/JPY with GBP/USD logarithmic return exchange rates (the bottom panel).
The graph illustrates the cross-correlation coefficient variation over scales $s$ ranging from 5 minutes up to 2 weeks
for $q=1, 2, 3, 4$.}
\label{pqs}
\end{figure}

As a more quantitative extension of this analysis we consider therefore some examples of the detrended cross--correlation $q$--coefficient $\rho_q(s)$ results for exchange rate series.
Two examples of the cross--correlation between two series of returns for exchange rates are shown in Fig.~\ref{pqs}: the top panel presents a pair of EUR/JPY and GBP/JPY,
whereas the bottom panel illustrates the results of $\rho_q(s)$ for EUR/JPY and GBP/USD.

>From the data shown in Fig.~\ref{pqs} it follows that the return rates for exchange rates EUR/JPY are markedly more correlated with the returns for exchange rates GBP/JPY than with the GBP/USD rate. 
This seems to be expected as in the former case there is a common (base) currency (JPY). In such a way a pair of returns is intrinsically correlated by JPY currency performance due to the triangular constraint
in the exchange rates.  This is an example of cross--correlations among 3 currencies. In this case the cross--correlations are in the triangular relation (the top panel of Fig.~\ref{pqs}). 
In the case shown in the bottom panel of Fig.~\ref{pqs} illustrating a pair of returns for exchange rates EUR/JPY vs GBP/USD which includes 4 different currencies, there is no triangular relation among them. 

However for both pairs of exchange rates (one of them pertains to the triangular relation, whereas the other one does not in the shown example) we observe that the larger size fluctuations are considered 
(the greater $q$ value is taken), the smaller cross--correlation in terms of the detrended cross--correlation $q$--coefficient $\rho_q(s)$ is observed over the time scale considered.
The magnitude of this cross--correlation measure is weakly dependent on the time scale and only slightly grows with time. Its growth is more pronounced for larger fluctuations (cf. data for $q=4$) than
this is the case for smaller fluctuations (cf. $q=1$).

\subsection{Fluctuations of different magnitude and cross--correlations}

\begin{figure}[h!]
\centering
\includegraphics[scale=0.42,keepaspectratio=true]{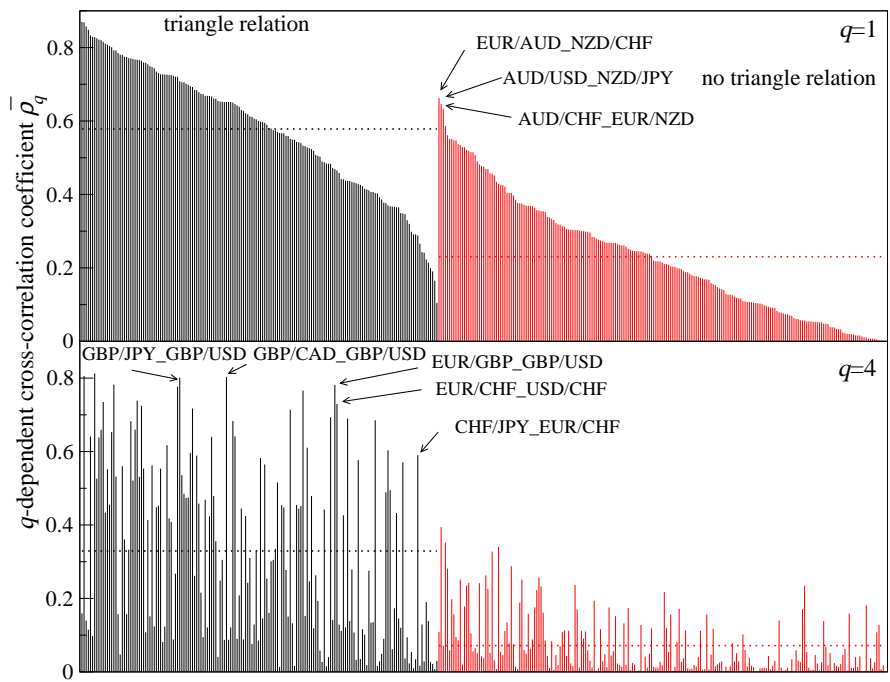} 
\caption{(Color online) Absolute values of cross--correlation coefficients $\rho_q$ for all possible pairs of currency exchange rates in the case of $q=1$ and $q=4$ averaged over all time scales $s$ ranging from $5 \  \textrm{mins}$ up to $2 \ \textrm{weeks}$. On the top--left and top--right panels the results for  $\rho_q$ are sorted in decreasing order. On the bottom--left and bottom--right panels the results are unsorted making an easier task to identify particularly high cross-correlations (shown with labels) for each of the both cases, the triangular and non--triangular relationship.}
\label{pqs14}
\end{figure}

Let us investigate in some more details the cross--correlations between relatively small and large fluctuations of two exchange rate return series. The results for all $\rho_q(s)$ cross--correlations (378 pairs in total)
over the whole period 2010--2018 in the case of $q=1$ and $q=4$ averaged over all scales $s$ are shown in Fig.~\ref{pqs14}. Note that the cross--correlation pairs are grouped into two classes. 
One class of exchange rate pairs (in black, left top and bottom panels) which pertain to the triangular relation and the second class, where cross--correlated pairs are outside the triangular relation
(in red, right top and bottom panels). On the top--left (in black) and top--right (in red) panels, the results for  $\rho_q(s)$ with $q=1$ are sorted in decreasing order. 
This ordering is unchanged for the purpose of the presentation of results with $q=4$. 

This gives an idea about the range of obtained values of cross--correlation coefficient distributions for the currency pairs
which are in or out the triangular relation. One can easily verify that at the level of smaller fluctuations ($q=1$) many pairs of logarithmic exchange rate returns, which are not linked by the triangular relation,
could have a higher averaged cross--correlation coefficient than many pairs which have such property. The black dotted horizontal line on the top--left panel shows the average cross--correlation of different pairs
pertaining to the triangular relation. The value of that overall average is about 0.6.

Note, that in the case of cross--correlations with no triangular relations between pairs of currencies including AUD and NZD we observe stronger correlations than in the case of pairs pertaining to the triangular relations 
(the top panel for $q=1$). It indicates a possibility of observing stronger correlations in exchange rates among four currencies in comparison with what we would expect on average in the case of exchange rates
linked with the triangular relations. This somewhat unexpected result could be ascribed to mechanisms coupling economies of these two countries. 

Hence, from our study it follows, that indeed some cross--correlations of the pairs, which are not linked by a common currency and are traded on the Forex market, may reach that overall average cross--correlation of exchange
rates with a common base. This seems to be a surprising conclusion, since typically we would expect stronger correlations between explicitly correlated two series (by means of a common, base currency) rather than in a case
where there is no such common base. However we have to appreciate the fact, that cross--correlations between any pair of exchange rates will have some impact on the cross--correlations of other pairs through mutual 
connections arising from different combinations of currencies being exchanged. Nonetheless, the small fluctuations in logarithmic returns would be difficult to use in viable trading strategies, mainly due to finite 
spreads in bid and ask rates. From the practical point of view, correlations of large fluctuations seem to be more
promising in finding and exploiting arbitrage opportunities.
The bottom panels show the results for the large fluctuations ($q=4$) in the corresponding cases of the exchange currency pairs which are in or outside the triangular relation (that is with a common currency or not). 
The order of currency pairs in the bottom panels is kept the same as in the top panels. For the case of the larger fluctuations, the level of overall average of cross--correlations is marked again with horizontal black
dotted line at a value of 0.3 approximately for the case of triangular--related class of currency pairs. The cross--correlations of the large fluctuations are therefore approximately two times smaller than in the case
of small correlations. However, in the case of GBP and CHF taken as the common base currency the cross-correlations are stronger for $q=4$ than for $q=1$. In the case of cross--correlations outside the triangular relations,
the strong cross--correlations arise when we take AUD as base currency on the one side and NZD on the other. 

It can be noticed that, in general the cross--correlations for smaller fluctuations ($q=1$) are stronger than cross--cor\-re\-la\-tions for the large fluctuations ($q=4$). This is in agreement with previous
findings~ \cite{Kwapien2015,Zhao2018,Watorek2018} for other financial instruments like stocks and commodities. The time evolution of averaged cross--correlations over currency pairs with the common base 
for large fluctuations will be discussed later (cf. Fig.~\ref{avecorr}).

\subsection{Dendrograms --- agglomerative hierarchical trees from cross--correlations of exchange rates}

As we have already seen, studying quantitative levels of \hfill\break 
cross--cor\-rela\-tions may uncover some less obvious con\-nec\-tions among currencies than just the explicit link through a common base currency.
In order to uncover a hierarchy of currencies in terms of logarithmic returns from exchange rates, one may consider cross--correlation coefficients as a measure of the distance $d(i,j)$ between different exchange rate pairs.

\begin{figure}
\includegraphics[scale=0.32,keepaspectratio=true,angle=0]{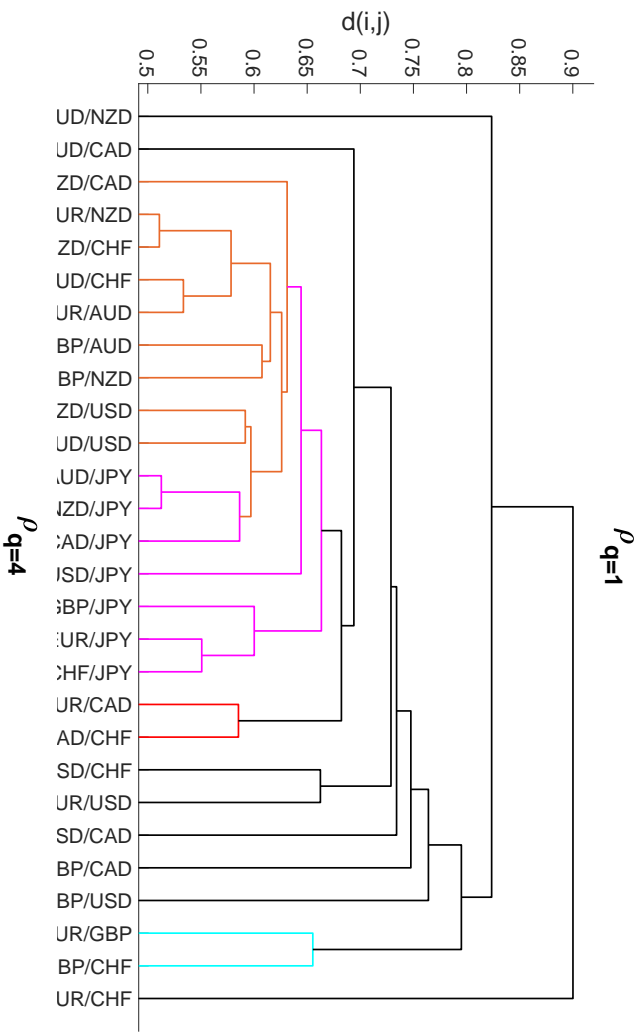} 
\\
\includegraphics[scale=0.32,keepaspectratio=true,angle=0]{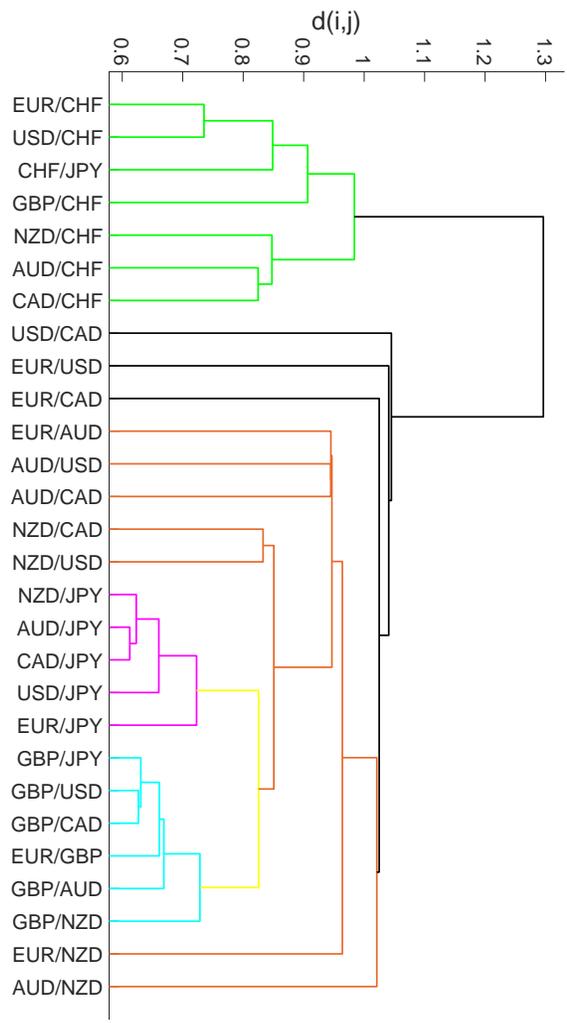}
\caption{(Color online) Dendrograms corresponding to  Fig.~\ref{pqs14} based on correlation matrices for $q=1$ and $q=4$ and averaged over time scales $s$.}
\label{dend14a}
\end{figure}

We define the distance $d(i,j)$ similarly to the definition introduced by \cite{Mantegna1999}, but instead of correlation coefficient, we use $q$--dependent cross--correlation coefficient~\cite{Kwapien2015,Kwapien2017}.
Thus the distance for agglomerative hierarchical trees take the following form:

\begin{equation}
d(i,j) = \sqrt{2 \Bigl ( 1 - \rho_q(s)(i,j) \Bigr )}
\label{dij} 
\end{equation}

It is worth noting, that to the best of our knowledge, it is the first ever such use of the cross--correlation $q$--coefficient as a way to induce measure for creating a hierarchy tree (a dendrogram).
As a result of adopting the distance given by Eq.~(\ref{dij}), we obtain graphs for the small ($q=1$) and for the large fluctuations ($q=4$) as it is shown in Fig~\ref{dend14a}.

For $q=4$ (the bottom panel of Fig~\ref{dend14a}, the cluster structure is more pronounced than it is in the case of $q=1$. In the $q=4$ case a well separated cluster structure is formed 
with CHF (green), GBP (cyan), JPY (magenta), AUD and NZD (orange) while in $q=1$ case we see mostly JPY (mangeta) and AUD/NZD (orange) with not so much hierarchical separation among various groups of currencies.
An interesting observation follows that Australlian (AUD) and New Zealand (NZD) dollars are strongly correlated -- they appear together in the same clusters of exchange rates both for the small and large fluctuations.
This indicates a possibility of building strong cross--correlations between exchange rate pairs which do not have the same common base.
However if two different currencies are strongly correlated than they effectively behave like a common base and the appropriate cross--correlations may reveal themselves as an unexpectedly large value of the $\rho_q$ coefficient. 
Such findings are important when designing the trading strategies, both for optimizing portfolio and for its hedging. 

We would like to stress the fact that our method is not limited only to time series from the Forex and it may well be applied to the signals in a form of time series arising in other fields of research and applications.

\subsection{Cross--corelations in multiple time scales}

\begin{figure}
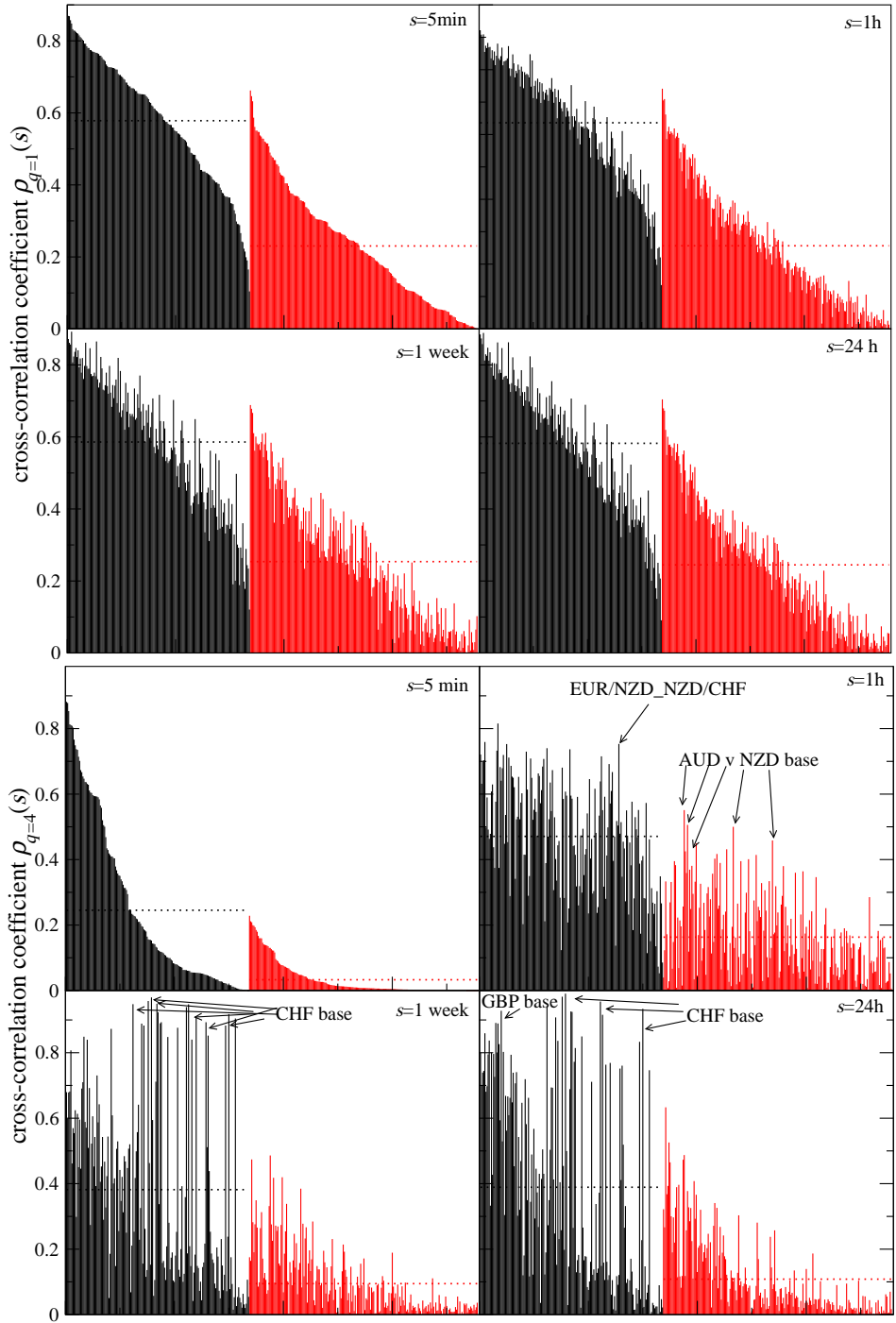

\centering
\includegraphics[scale=0.45,keepaspectratio=true]{6_analysis2010_18pos_q1.eps}
\includegraphics[scale=0.45,keepaspectratio=true]{6_analysis2010_18pos_q4.eps} 
\caption{(Color online) Detrended cross--correlations  $\rho_q(s)$ for all currency exchange pars (378 pairs in total) in the case of $q=1$ (top 4 panels) 
and $q=4$ (bottom 4 panels). For each $q$ value we consider 4 values of $s$ corresponding to $5 \ \textrm{mins}$, $1 \ \textrm{hour}$, $24 \ \textrm{hours}$ and $1 \ \textrm{week}$.}
\label{pqs14s}
\end{figure}

We have looked already into the cross--correlations within fluctuation magnitude domain. Let us now investigate the cross--correlations in the time domain. For that reason we compute the detrended cross--correlation
$q$--coefficient $\rho_q(s)$ in the case of four different time scales: $s = 5 \ \textrm{mins}$, $s=1 \ \textrm{hour}$, $s=24 \ \textrm{hours}$ and $s=1 \ \textrm{week}$ for the case of $q=1$ and $q=4$.
The results are shown in Fig.~\ref{pqs14s}. The top four panels refer to the $q=1$ case whereas the bottom four panels illustrate the case of $q=4$. In both cases, the cross--correlations are sorted in decreasing 
order of values obtained for the shortest time scale ($s = 5 \ \textrm{min}$). We sort $\rho_q(s)$ values for the shortest time scale and keep that ordering unchanged when taking longer time scales.

We also show the overall average for the currency exchange rate pairs complying to the triangular relation (the black dotted line) and for currency exchange rate pairs which are not bounded by
the triangular relation (red dotted line). The most striking feature when comparing the small and large fluctuation cross--correlations over different time scales, is that in the former case 
little is happening over different time scales considered. The plots indicate nearly static cross--correlations, almost independent on the time scale for the small fluctuations. 
On the other hand, the cross--correlations for large fluctuations ($q=4$) change with extension of the time scale $s$. 

The magnitude of values $\rho_q(s)$ for currency exchange rates which are in triangular relation (black/left part of each panel) stay within the same range although the values themselves fluctuate significantly with a change of time scale. 
Specifically, the overall average (denoted by the black dotted horizontal line) grows from a value which is less than 0.3 (for $s = 5 \ \textrm{mins})$) up to approximately 0.4 (for $s \ge 1 \ \textrm{hour})$.
The growth of the overall average of cross--correlation is even more convincing for the class of pairs of currency exchange rates which are not in a strict triangular relation.

In general for $q=4$ and longer time scales $s$ (24 hours, 1 week) we observe stronger cross--correlations within exchange rates, related by the triangular relationship, for the pairs where 
the Swiss franc (CHF) or the British pound sterling (GBP) is the common base currency. It is evident that the cross---correlations (red/right parts of each panel) obtained for $s = 5 \ \textrm{mins}$ 
are much smaller than those corresponding to  $s \ge 1 \ \textrm{hour}$.

What is more, for the shortest time scale shown here, the difference between cross--correlations for pairs that are in the triangular relations
and those that are not, is the biggest. The cross--correlation for currency exchange pairs outside the triangular relation in the case of large fluctuations in logarithmic return rates grows in time,
which indicates propagation of correlations in time. This explains why averaged cross--correlations for such currency pairs may be unexpectedly high (cf. Fig.~\ref{pqs14}).
This gives us some idea about the information propagation time through the Forex market, which is the time needed to reflect the maximum average cross--correlation between any pair of exchange currency rates. 
As we have already mentioned above, in the Forex market all currency rates are connected through mutual exchange rate mechanism. However in some cases the inherently stronger correlations 
(e.g induced by the triangular relationship) will result in a very fast response (so the cross--correlations are significant after a relatively short time) whereas for some other exchange rates such response 
would be lagged in time (e.g. carried over through a sequence of exchange rate adjustments). This time lag could be regarded as an estimate for the time duration of window of opportunity to execute an arbitrage opportunity.

\begin{figure}[h!]
\centering
\includegraphics[scale=0.44,keepaspectratio=true]{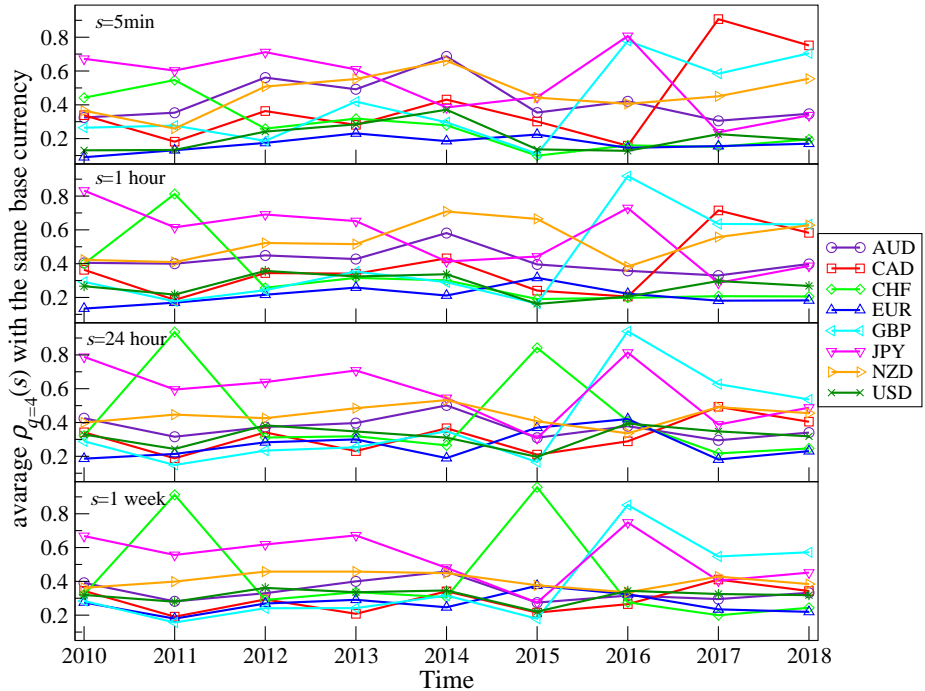}
\caption{(Color online) Average detrended cross--correlations $\rho_q(s)$ for $q=4$ and various scales $s$ with the same base currency
computed for each year separately. Note pronounced maxima related to CHF as a base currency (2011, 2015), GBP and JPY (2016).}
\label{avecorr}
\end{figure}

The results presented in Fig~\ref{avecorr} show how these averages change in consecutive years. Again we see clearly an increase for CHF taken as a base currency in 2011 and 2015 (SNB interventions).
The result is consistent for a range of time scales $s$ taken in our approach. It is interesting to note, that this effect becomes more pronounced when longer time scales are implemented (24 hours, 1 week).
A similar conclusion is valid when considering GBP or JPY taken as the base currency -- corresponding curves have a maximum in 2016. GBP was mainly influenced by Brexit and GBP flash crash, whereas JPY appreciation
was caused indirectly by US elections. On the contrary, the increase in the average for CAD as the base can be noted only for the shortest time scale $s= \ 5 \textrm{min}$ in 2017.
This could mean that the sudden overnight increase of the rates by the Bank of Canada in 2017 did not have longer lasting effect and was only causing very short term effect. 
In order to identify promising arbitrage opportunities (e.g. the triangular arbitrage) our tentative conclusion would be that we have to observe large fluctuations 
(by means of significant cross--correlation coefficients for $q=4$) for relatively long time scale $s$.

In view of the above findings where we have already identified an important role of the large fluctuations,
a question arises to what extent (even briefly occurring in time) such extreme events (fluctuations) in currency exchange returns may influence the detrended cross--correlations.
In the following let us investigate the impact of extreme events on the cross--correlations expressed by the $\rho_q(s)$ coefficient. In Fig.~\ref{fx} we have seen possible influence of various events on the currency index.
It is interesting to see how these extreme events manifest themselves as far as cross--correlations are concerned.

\begin{figure}[h!]
\centering
\includegraphics[scale=0.45,keepaspectratio=true]{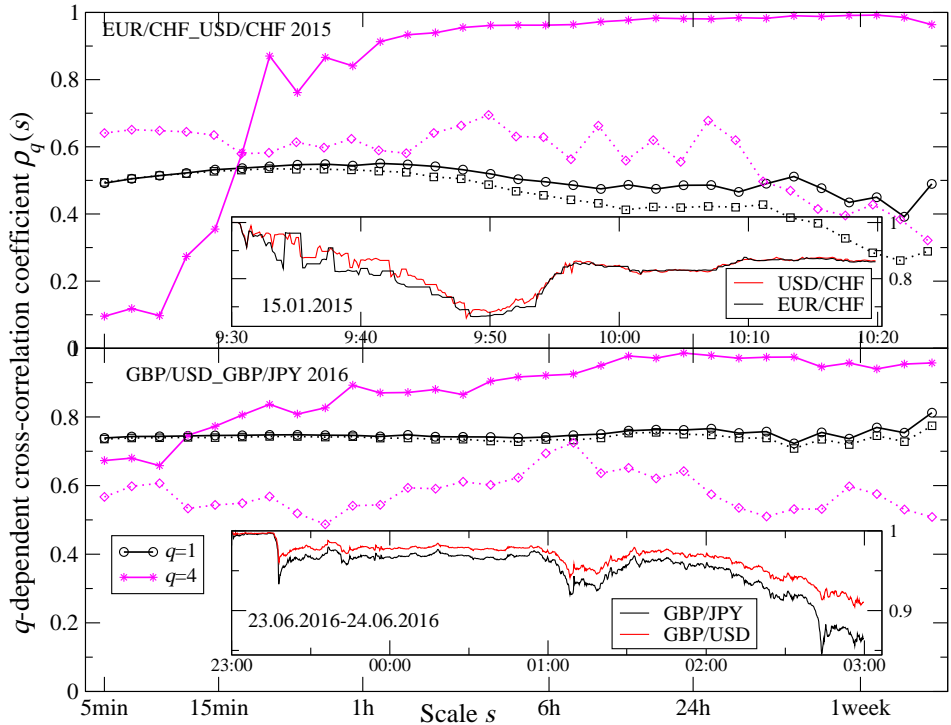}
\caption{(Color online) Example cross--correlations EUR/CHF with USD/CHF in 2015 and GBP/USD with GBP/JPY in 2016 with $q=1$ and $q=4$. These exchange rates exhibit substantial volatility during considered years. The dashed line corresponds to the cross--correlation results with rejected periods of time with large volatility and existence of triangular arbitrage opportunities. Insets show cumulative log returns during the SNB intervention with CHF (2015) exchange rates and during the night just after Brexit referendum (GBP exchange rates).}
\label{pqb_noextreme}
\end{figure}

The periods of extreme variation of exchange rates are shown in the corresponding insets of Fig.~\ref{pqb_noextreme}. In the case of the small fluctuations ($q=1$, black solid and dotted curves),
these periods of significant exchange rate variations do not have significant impact on overall behavior through a range of time scales $s$. However the large fluctuations 
in the cross--correlations ($q=4$, magenta solid and dotted curves) are highly sensitive to the existence of these well localized in time events. 
The insets show that in fact the exchange rates compared (red and black curves) were changing so rapidly that they could not follow each other. In such a way the possible arbitrage opportunities have arisen.

\subsection{Detecting triangular arbitrage opportunities}

\begin{figure}[h!]
\centering
\includegraphics[scale=0.45,keepaspectratio=true]{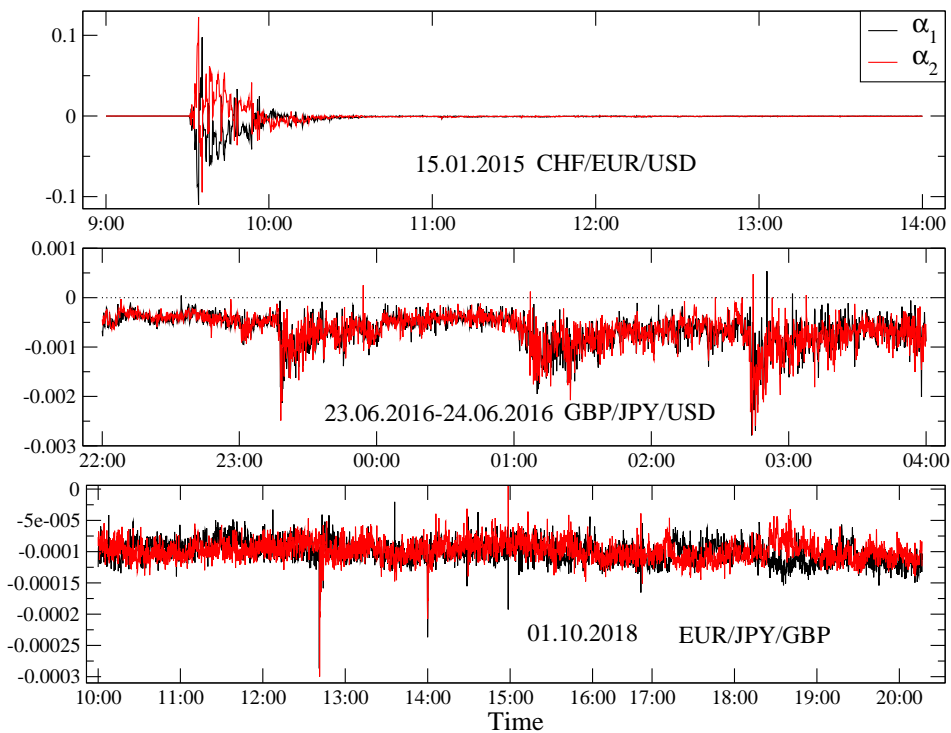}
\caption{(Color online) Deviations from  the triangular relations. In 2015 existed a big arbitrage opportunity (CHF), moderate arbitrage opportunity (GBP) in 2016 and no such opportunity in 2018. In this case we use ask and bid prices for exchange rates (instead of averaged ones) in order to show this in more details.}
\label{opport}
\end{figure}

Finally let us investigate closely these brief in time periods of arbitrage opportunities we have identified by our data analysis. Fig.~\ref{opport} shows 
triangular arbitrage coefficient $\alpha_1$ and $\alpha_2$ (cf. Eqs.(\ref{a1}), (\ref{a2})).

All events indicated by values greater than 0 in fact could potentially offer triangular arbitrage opportunities. 
The top panel shows an example of potentially significant arbitrage opportunity which is related to the SNB intervention in 2015 and fluctuations in the CHF exchange rates. 
The middle panel of Fig.~\ref{opport} illustrates a moderate arbitrage opportunity, which is related to the Brexit referendum and its impact on the GBP exchange rates. 
Finally, the bottom panel illustrates rather weak chance of exploiting triangular arbitrage opportunity --- there is only one very brief in time instance when in theory this might be possible.
In such a way we have shown that the inspection of cross--correlations by means of the $\rho_q(s)$ coefficient over a range of time scales $s$, especially for the large fluctuations 
(in our multifractal approach with the maximum value of the parameter $q=4$ to ensure the convergence) allows to uncover the hierarchy in exchange rates and focus on the periods of time 
when arbitrage opportunities potentially exist. The arbitrage opportunities are very closely related to large fluctuations which tend to be more pronounced in the longer time scales $s$.
This is the case for exchange rates related to CHF and GBP and this is precisely what opens windows of opportunities for the triangular arbitrage.

\section{Conclusions}

We have investigated currency exchange rates cross-correla\-tions within the basket of 8 major currencies. Distributions of 10--second historical logarithmic exchange rate returns follow approximately
the inverse cubic power--law behavior when the brief period of trading on the 15th of January 2015 in the wake of the SNB intervention is excluded from the exchange rate data sets.
The tails of the cumulative distributions of the high--frequency intra--day quotes exhibit non--Gaussian distribution of the rare events by means of the so called ``fat tails'' (large fluctuations). 
This clearly documents that large fluctuations in the logarithmic rate returns occur more frequently than one may expect from the Gaussian distribution. 

We have found that on average the cross--correlations of exchange rates for currencies in the triangular relationship are stronger than cross--correlations between exchange rates for currencies outside
the triangular relationship. Detrended cross--correlation approach with $\rho_q$ cross--correlation measure allows to uncover a hierarchical structure of exchange rates among set of currencies 
in the Forex market. To the best of our knowledge, it is the first ever application of the $\rho_q$ cross--correlation coefficient to the agglomerative hierarchical clustering in a form of dendrograms.  
Such dendrograms may have important applications related to hedging, risk optimization and diversification of the currency portfolio in the Forex market.

The detrended cross--correlation coefficient $\rho_q$ is sensitive to time scales and fluctuation magnitudes allowing for a subtle (more sophisticated) cross--correlation measuring than 
just a simple global estimates (e.g. the Pearson's coefficient). By using $\rho_q$ coefficients one may discriminate subintervals of time series when significant changes in time of cross--correlations 
are observed. Such abrupt changes of cross--correlations combined with the presence of relatively large fluctuations may signal potential triangular arbitrage opportunities.
We have shown a viable possibility of such application using historical data for years  2015 and 2016, when a significant enhancement of correlations in exchange rate pairs
involving CHF, GBP, JPY  for large fluctuations ($q=4$) has been observed with increasing the time scale $s$.

Finally, our conjecture is that during significant events (e.g. with an impact on the Forex market) one may expect existence of potential triangular arbitrage opportunities. 
Such events and opportunities we have indeed identified in the historical trading data of the years 2010--2018. 
The evidence we have shown clearly indicates that the multifractal cross--correlation methodology should contribute significantly to predictive modelling of temporal
and multiscale patterns in time series analysis.

We believe that our present study, where we consider currencies interaction through their mutual exchange rates and the dynamics of the rates adjustment to a new conditions due to a sudden event, may encourage future research in studying the information propagation through complex networks of interacting entities. This in turn may have some consequences for design of new smart learning methods for neural networks and a general computational intelligence in predicting a future behavior of complex systems. For example, since we have demonstrated feasibility of financial time series analysis against favorable patterns,  we may expect future advancement in computer algorithms for financial engineering when trading tick--by--tick data are available in real time.


\section*{Appendix}

The data used in the present study has been obtained from the Dukascopy Swiss Banking Group \cite{data}.
The data set comprises foreign exchange high--frequency bid and ask prices for the period 2010 -- 2018 with the time interval $\Delta t = 10 \ \textrm{s}$. We consider exchange rates between pairs of the following 8 major currencies:
Australian dollar (AUD), Canadian dollar (CAD), Swiss franc (CHF), euro (EUR), British pound sterling (GBP), Japanese yen (JPY), New Zealand dollar (NZD) and US dollar (USD). 

Thus we have in the data set all 28 exchange rates (bid and ask prices) among the set of 8 currencies. 
Hence the foreign exchange rates are the following: \\
AUD/CAD, \ AUD/CHF, \ AUD/JPY, \ AUD/NZD, \ AUD/USD, \ CAD/CHF, 
CAD/JPY, \ CHF/JPY, \ EUR/AUD, \ EUR/CAD, \ EUR/CHF, \ EUR/GBP,  
EUR/JPY, \ EUR/NZD, \ EUR/USD, \ GBP/AUD, \ GBP/CAD, \ GBP/CHF, 
GBP/JPY, \ GBP/NZD, \ GBP/USD, \ NZD/CAD, \ NZD/CHF, \ NZD/JPY, 
NZD/USD, \ USD/CAD, \ USD/CHF, \ USD/JPY.

The quotes have the meaning of {\it indicative} bid/ask prices rather than {\it executable} prices. The indicative and executable prices differ typically by a few basis points \cite{Aiba2004,Fenn2009}.
Bid/ask (sell/buy) quotes give the spread, that is a difference in prices at which one can buy/sell a currency. The ask price is greater than or equal to the bid price. The spread is dependent on the liquidity 
(a number and volume of transactions) as well as on some other factors.

We filter out such raw data (time series) by removing periods when for any given pair there was no quote available or no trading (e.g. weekends, holidays), which effectively resulted in the rate being unchanged.
Typically, we thus have approximately 2.2 million data points per year and therefore this amounts to approximately 20 million of observations for each exchange rate time series.

We have used MATLAB by MathWorks numerical computing environment for implementing our time series methodology with the numerical analysis performed on a modern PC class computer. 
Validation methods included a simple model parameter variation as well as surrogate data methods. We can apply our procedure to randomly shuffled original data.
We can also create Fourier surrogate time series. In the latter validation method, the Fourier transform of the original time series is computed and then 
the inverse Fourier transform is applied to the retained amplitudes, but randomly mixed phases \cite{Kwapien2012,Kwapien2015}.

\newpage
{\footnotesize

\end{document}